\documentclass[12pt,halfline,a4paper]{ouparticle}

\usepackage[utf8]{inputenc} % fallowutf-8 input
\usepackage[T1]{fontenc}    % use 8-bit T1 fonts
\usepackage{hyperref}       % hyperlinks
\usepackage{url}            % simple URL typesetting
\usepackage{booktabs}       % professional-quality tables
\usepackage{amsfonts}       % blackboard math symbols
\usepackage{nicefrac}       % compact symbols for 1/2, etc.
\usepackage{microtype}      % microtypography
\usepackage{lipsum}
\usepackage{fancyhdr}       % header
\usepackage{amsfonts, amssymb, amsmath, amsthm, mathtools, bm}
\usepackage{graphicx}       % graphics
\graphicspath{{media/}}     % organize your images and other figures under media/ folder
\usepackage{derivative}
\usepackage[dvipsnames]{xcolor}

\usepackage[style=authoryear, backend=biber, sorting=nyt, maxbibnames=5, maxcitenames=2, natbib]{biblatex}
\bibliography{references.bib}

\DeclareMathOperator{\Tr}{Tr}

\begin{document}

\title{Efficient smoothness selection for nonparametric Markov-switching models via quasi restricted maximum likelihood}

\author{%
\name{Jan-Ole Koslik}
\address{Bielefeld University, Department of Business Administration and Economics, Bielefeld, 33615, Germany}
\email{jan-ole.koslik@uni-bielefeld.de}
\and
% \name{Second Author}
% \address{Institute or Organization, Department, City, State,\\
% Zip Code, Country}
% \email{e-mail address}
}

% \abstract{
% We introduce a novel approach for smoothness selection in nonparametric Markov-switching models via restricted maximum likelihood estimation (REML), drastically reducing the computational burden compared to other existing methods. Traditional techniques, such as K-fold cross-validation, and information criteria-based model selection, suffer from significant drawbacks, including difficulty in achieving a suitable bias-variance trade-off and the reliance on computationally expensive grid search methods. These limitations become particularly pronounced with an increasing number of smooth effects, hampering practical usability. While \citet{michelot2022hmmtmb} also proposed smoothness selection for Markov-switching models via restricted maximum likelihood estimation based on the TMB package, this approach results in a nested optimization problem, requiring considerable computational resources. We apply a mathematical trick allowing for a substantial reduction in the computational burden while yielding higher accuracy regarding the estimation of fixed effects parameters. Hence, the proposed REML method provides a more efficient and reliable mechanism for smoothness selection, making it feasible to apply P-splines in practical applications involving rich and intricate data structures, paving the way for broader and more effective utilization of nonparametric methods in diverse fields such as ecology, finance, and medicine.
% }

\abstract{Markov-switching models are powerful tools that allow capturing complex patterns from time series data driven by latent states.
Recent work has highlighted the benefits of estimating components of these models nonparametrically, enhancing their flexibility and reducing biases, which in turn can improve state decoding, forecasting, and overall inference.
%thereby increasing flexibility and decreasing potential biases, which can improve state decoding, forecasts, and general inference.
Formulating such models using penalized splines is straightforward, but practically feasible methods for a data-driven smoothness selection in these models are still lacking.
Traditional techniques, such as cross-validation and information criteria-based selection suffer from major drawbacks,
%including difficulty in achieving a suitable bias-variance trade-off and the 
most importantly their reliance on computationally expensive grid search methods, hampering practical usability for Markov-switching models. \citet{michelot2022hmmtmb} suggested treating spline coefficients as random effects with a multivariate normal distribution and using the R package \texttt{TMB} \citep{kristensen2016tmb} for marginal likelihood maximization. While this method avoids grid search and typically results in adequate smoothness selection, it entails a nested optimization problem, thus being computationally demanding.
We propose to exploit the simple structure of penalized splines treated as random effects, thereby greatly reducing the computational burden while potentially improving fixed effects parameter estimation accuracy. 
%This approach, known as penalized quasi-likelihood (\textcolor{orange}{qREML}), has been used for generalized additive models (GAMs).
Our proposed method offers a reliable and efficient mechanism for smoothness selection, rendering the estimation of Markov-switching models involving penalized splines feasible for complex data structures.
% and enhancing the application of nonparametric methods to fields like ecology, finance, and medicine.
}

\date{\today}

\keywords{Markov-switching regression; regime switching; hidden Markov models; penalized splines; smoothness selection}

\maketitle

\section{Introduction}

Markov-switching models, also known as hidden Markov models (HMMs) and regime-switching models, are a versatile statistical tool for modeling time series data and have been applied successfully in a wide array of fields, including ecology \citep{mcclintock2020uncovering}, finance \citep{zhang2019high}, medicine \citep{amoros2019} and seismology \citep{beyreuther2008}. Their ability to model systems driven by underlying, sequentially correlated states makes them particularly valuable for understanding complex processes that exhibit regime shifts. However, the traditional formulation of HMMs often falls short in flexibility, failing to capture the intricate patterns and dependencies present in real-world data like non-linear covariate effects or non-standard emission distributions. Choosing adequate parametric emission distributions is particularly difficult, as the state-dependent empirical distributions cannot be visualized before fitting the model due to the underlying states not being known a priori \citep{langrock2018spline}. Over the last decade, these limitations catalyzed the development and application of nonparametric methods, which offer a more adaptable modeling framework, hence avoiding potential biases introduced by assuming parametric relationships.

The incorporation of nonparametric approaches into HMMs has been advanced by, \textit{inter alia}, \citet{dannemann2012semiparametric, langrock2015nonparametric, langrock2017markov, feldmann2023flexible, chen2023bayesian}, who demonstrated that the relaxation of parametric assumption by using \textit{penalized splines} can lead to more reliable inference and a more detailed description of real-world phenomena, allowing models to adapt to the data more naturally. Such models for example permit nonparametric emission distributions or smooth covariate effects on the state or state-dependent process. Despite all the work focussed on incorporating nonparametric aspects into HMMs, the most crucial part of model estimation, namely adequate smoothness selection for penalized splines, remains immensely challenging and has not been addressed satisfactorily. 
Traditional methods for smoothness selection like subjective choosing of the penalty parameter based on visual inspection, $K$-fold cross-validation (CV), or model selection based on information criteria like AIC and BIC 
% that account for the decrease in the degrees of freedom imposed by the penalty, 
are commonly used to date \citep{langrock2018spline} but come with considerable drawbacks. Most prominently, these methods rely on grid search techniques, resulting in an intense computational burden that increases exponentially with the number of smooth effects present in the model while severely limiting the practical potential of penalized splines for practical applications with potentially rich and complex models.

Recently, the use of marginal maximum likelihood within a random effects framework has emerged as a promising alternative \citep{michelot2022hmmtmb}. By treating the spline coefficients as random effects that are integrated out during model estimation, this approach tends to produce parsimonious models, effectively balancing complexity and fit --- a fact known for a long time for generalized additive models (GAMs) \citep{reiss2009smoothing}. 
The approach relies on the \texttt{TMB} R package \citep{kristensen2016tmb} which employs the so-called \textit{Laplace approximation} to compute the high-dimensional integral. In turn, this leads to a nested optimization problem where for each Newton-Raphson step in the fixed effects and penalty strength optimization, an inner optimization over the spline coefficients needs to be conducted. By leveraging automatic differentiation (AD), this approach represents a significant advancement, making nonparametric modeling both reliable and practical for HMMs for the first time. However, the nested optimization leads to an additional layer of complexity and computational burden to the estimation process that can be challenging to manage, specifically as parameter estimation for HMMs can be numerically challenging even for rather simple parametric models. 

In this contribution, we argue that for the special case of penalized splines formulated as random effects, the intense computational overhead implied by the nested optimization is indeed not necessary. We demonstrate that ideas originally proposed for GLMMs by \citet{laird1982random, schall1991estimation}
% \citet{breslow1993approximate, wood2011fast} 
can be applied to HMMs, drastically reducing the computational burden of marginal ML by converting the nested (numerical) optimization problem to a procedure that ultimately results in iterating model fitting based on sequentially updated penalty strengths. This not only substantially decreases the computational overhead, thus speeding up model estimation, but additionally allows for an exact estimation of the fixed effects parameters, which are influenced by the quality of the Laplace approximation only indirectly through the chosen penalty strength. Furthermore, we implement the proposed method as an easy-to-use R function, making model fitting easy for anyone familiar with direct numerical maximum likelihood estimation. We thus believe this method to represent a crucial step towards the establishment of nonparametric Markov-switching models in applied time series analysis.

The paper proceeds as follows. Section \ref{sec:model formulation}, introduces the basic HMM model formulation and reviews three popular options of including nonparametric elements in different model aspects. Subsequently, Section \ref{subsec:PML} discusses parameter estimation using penalized maximum likelihood estimation for a given penalty strength parameter. Building on this foundation and existing approaches for smoothness selection detailed in Section \ref{subsec:existing}, Section \ref{subsec:qREML} provides a novel approach for estimating the penalty strength based on the random effects framework and inspired by smoothness selection for GAMs. 
%Following this, Section \ref{sec:simulation} demonstrates the proposed approach through simulation experiments, and finally,
Finally, Section \ref{sec:case studies} showcases three case studies, each corresponding to one of the discussed modeling options. Additional numerical experiments are described in Appendix \ref{sec:simulation}.

\section{Model formulations}

\label{sec:model formulation}

\subsection{Basic model formulation}

A basic Markov-switching model or HMM is a doubly stochastic process, comprising an observed process $\{X_t\}_{t = 1, \dotsc, T}$ and a latent, unobserved state process $\{S_t\}_{t = 1, \dotsc, T}$, taking values in $\{ 1,\ldots, N\}$. Conditional on the value of $S_t$, the observation $X_t$ is independent of all previous and future values of both the observed process and the state process, which is formally known as the \textit{conditional independence assumption}. Thus, the conditional distribution of $X_t$ is fully specified by the family
$$
f_i(x_t) = f(x_t \mid S_t = i), \quad i = 1, \dotsc, N,
$$
where $f$ denotes either a probability density function if $X_t$ is continuous or a probability mass function if $X_t$ is discrete. Naturally, the $f_i$ are called state-dependent (or emission) distributions, and hence $X_t$ is also called the state-dependent process. In cases where $X_t$ is vector-valued, the $f_i$ are multivariate which is often achieved by assuming independence of the components of $X_t$ given the underlying state --- rendering $f_i$ the product of several univariate distributions --- or by using an appropriate multivariate distribution such as a multivariate normal.

Lastly, the state process $\{S_t\}$ is assumed to be a Markov chain of first-order, fully characterized by its initial state distribution $\bm{\delta}^{(1)}$ and the possibly time-varying transition probability matrix (t.p.m.)
%$$\boldsymbol{\Gamma}^{(t)}=(\gamma_{ij}^{(t)}), \; \text{ with } \; \gamma_{ij}^{(t)}=\\Pr(S_{t}=j | S_{t-i}=i),$$
$$\boldsymbol{\Gamma}^{(t)}=(\gamma_{ij}^{(t)}), \; \text{ with } \; \gamma_{ij}^{(t)}=\Pr(S_{t}=j \mid S_{t-1}=i), \quad t = 2, \dotsc, T.$$
Within this basic model formulation, one can and should investigate every aspect of the model regarding the assumptions made and, if necessary, allow for flexibilization thereof. For example, it may be unrealistic to assume state-dependent distributions of a particular parametric form, or polynomial covariate effects, 
%on either the transition probabilities or the state-dependent distributions
hence suggesting different ways of model flexibilization that allow for more realism. To achieve this, we focus on penalized splines, which are particularly attractive due to their mathematical simplicity and computational advantages, making them ideal tools to embed within larger model formulations \citep{langrock2018spline}.

Before discussing model estimation and smoothness selection very generally in Section \ref{sec:smoothness} --- i.e.\ without restriction to a particular HMM-type model formulation --- we first give a brief overview of three popular extensions of the basic model above to give a better intuition for use of splines within Markov-switching models.

% While Section \ref{sec:smoothness} discusses model estimation and smoothness selection very generally, for any kind of nonparametric relationship included in the model that is based on penalized splines, we start by giving a brief overview that outlines three popular extensions of the basic model formulation above, developed within the last years.

\subsection{Nonparametric state-dependent densities}

\label{subsec:nonparametric_densities}

The state-dependent distributions within an HMM are commonly assumed to belong to parametric families, thus being fully specified by a relatively small number of parameters. However, such an approach may be too restrictive to accurately model the observed process or may require selecting between many different candidate models. In particular, an inadequate choice of the parametric family of state-dependent distributions can lead to an overestimation of the number of states \citep{pohle2017selecting, langrock2015nonparametric}. Additionally, selecting appropriate parametric families for the state-dependent distributions is notoriously difficult, as the empirical distributions cannot be visualized a priori due to the latent states being unknown.

Flexible spline-based modeling of densities has been discussed by \citet{schellhase2012density} in general and by \citet{langrock2015nonparametric} for HMMs. Specifically, for univariate $X_t \in \mathbb{R}$, the state-dependent densities can be modeled by finite linear combinations of fixed basis functions $B_0, \dotsc, B_K$ as
$$
f_i(x) = \sum_{k=0}^K \alpha_k^{(i)} B_k(x), \quad i = 1, \dotsc, N.
$$
For $f_i$ to be a probability density function, the basis functions need to integrate to one and the coefficients $\alpha^{(i)}_{k}$ need to satisfy $\sum_{k=0}^K \alpha_k^{(i)} = 1$ and $\alpha_j^{(i)} \geq 0$ for all $k = 0, \dotsc, K$. The latter constraint is enforced by modeling the coefficients as transformations of unconstrained parameters $b_k^{(i)} \in \mathbb{R}$ using the multinomial logit link, i.e.\ $\alpha_k^{(i)} = \exp(b_k^{(i)})/\sum_{j=0}^K \exp(b_k^{(i)})$, setting one parameter $b_k^{(i)} = 0$ for identifiability.
% We set $b_0^{(i)} = 0$ to ensure identifiability, thus for each state $K$ parameters can vary freely. 
While any probability density function can be used for the basis expansion, B-spline basis functions with equally spaced knots in the domain of $X_t$ are a particularly convenient and numerically stable choice.
The number of basis functions should be chosen sufficiently large to allow for flexible shapes of the estimated density, such that any biases due to limited model flexibility can be avoided. Overfitting is then controlled by imposing a penalty on the curvature of $f_i$ (see Section \ref{subsec:PML}).
%Referring to \cite{} ... is typically sufficient.

Generalizing this approach to multidimensional settings is in principle straightforward by using tensor product B-splines (see \cite{michels2023nonparametric} and \cite[][Chapter 8]{fahrmeir2022regression}), but is only feasible in low dimensions due to the curse of dimensionality.
%It is of course possible to assume independence of the components of $X_t$ given the underlying state, but this seems inappropriate given that the scope of nonparametric modeling is to accurately represent the true distribution --- accounting for potentially complex dependencies between dimensions.

\subsection{Markov-switching generalized additive models}

\label{subsec:MSGAM}

Markov-switching generalized additive models (MS-GAMs) extend basic HMMs by allowing the expectation of the state-dependent distribution to depend on covariates $\bm{z}_t = (z_{t1}, \dotsc, z_{tQ})$. More formally, we denote the conditional density (or p.m.f.) of $X_t$ as $f(x_t ; \mu_t^{(i)}, \phi^{(i)})$ where for each state $i$, the mean follows the relationship
$$
g(\mu_t^{(i)}) = \eta^{(i)}_t = \beta_0^{(i)} + \sum_{q=1}^Q s_q^{(i)}(z_{tq}), \quad i = 1, \dotsc, N,
$$
and $\phi^{(i)}$ is a state-specific dispersion parameter, as described by \citet{kim2008estimation} and \citet{langrock2017markov}. Here, we omit possible linear effects of some covariates for notational simplicity. The model formulation above implies that each state of the Markov chain is linked to a specific GAM, with the transition dynamics determining which regression is active at any given time point (see \cite{wood2017generalized} for an exhaustive discussion of GAMs). Such models have been applied to financial as well as ecological data \citep{hambuckers2018markov, byrnes2023daily}.

The smooth functions $s_q^{(i)}$, $q = 1, \dotsc, Q, \; i = 1, \dotsc, N$ are conveniently expressed as finite linear combinations of fixed basis functions
$$
s_q^{(i)}(z) = \sum_{k=0}^K b_{qk}^{(i)} B_{qk}(z),
$$
where again, one coefficient $b_{qk}^{(i)}$ is set to zero for identifiability.
%the coefficients $b_{q0}^{(i)}$ are set to zero for identifiability.
%with cubic B-spline basis functions, among others, again being a popular option. 
%Choosing a relatively high number of basis functions for sufficient flexibility and penalizing the curvature of the smooth functions renders this option P-spline smoothing \citep{eilers1996flexible, fahrmeir2022regression}.

A straightforward generalization is to consider Markov-switching generalized additive models for location scale and shape (MS-GAMLSS), by linking additional parameters of the state-dependent distributions to covariates via additional smooth functions and suitable link functions (see \cite{adam2022gradient}).
% for a detailed discussion of MS-GAMLSS). In the most simple generalization, we could additionally model the state-dependent dispersion parameter as a smooth function of covariates --- as will be demonstrated in Section \ref{subsec: energy}.
%--- but it is also possible to choose a response distribution with more than two parameters.

\subsection{Nonparametric modeling of transition probabilities}

\label{subsec:transprobs}

Especially in behavioral ecology, it has become commonplace to model the transition probabilities of the underlying Markov chain as functions of covariates, with the aim of inferring internal and external drivers of animal behavior \citep{van2019classifying}.
%Letting $\bm{z}_t = (z_{t1}, \dotsc, z_{tq}) \in \mathbb{R}^q$ to be a set of covariates for $t = 1, \dotsc, T$. 
Conditional on the Markov chain being in state $i$ at time point $t-1$, the categorical distribution of states at time point $t$ is given by $(\gamma_{i1}^{(t)}, \dotsc, \gamma_{iN}^{(t)})$. Each of these categorical distributions, i.e.\ for $i = 1, \dotsc, N$, can now be modeled using a multinomial logistic regression
$$
\gamma_{ij}^{(t)} = \frac{\exp(\eta^{(ij)}_{t})}{\sum_{k=1}^N \exp(\eta^{(ik)}_{t})} \quad i,j = 1, \dotsc, N,
$$
where for $i \neq j$, $\eta^{(ij)}_{t}$ are predictors and $\eta^{(ii)}_{t}$ is set to zero for $i=1,\dotsc, N$ to ensure parameter identifiability. Similar to the previous section, the predictors are
% $$
% \eta^{(ij)}_{t} = \beta_0^{(ij)} + \beta_1^{(ij)} z_{t1} + \dotsc +  \beta_p^{(ij)} z_{tQ}  = \bm{\beta}^{(ij) \: '}\bm{z}_t,
% $$
% by adding an intercept column and $\eta^{(ii)}_{t}$ is set to zero for $i=1,\dotsc, N$ to ensure parameter identifiability. The natural extension to nonparametric modeling is to allow for an additive predictor of smooth functions
$$
\eta^{(ij)}_{t} = \beta^{(ij)}_0 + \sum_{q=1}^Q s^{(ij)}_q(z_{tq}),
$$
where again, we omit the possibility of linear effects of some covariates for notational simplicity. As above, the smooth functions can be represented by finite linear combinations of fixed basis functions. 
% $$
% s^{(ij)}_q(z) = \sum_{k=0}^K b_{qk}^{(ij)} B_{qk}^{(ij)}(z),
% $$
% with scaling coefficients $b_{q1}^{(ij)}, \dotsc, b_{qK}^{(ij)} \in \mathbb{R}$ to be estimated and again $b_{q0}^{(ij)}$ set to zero for identifiability. 
Again, many options exist for the choice of basis, with cubic B-spline basis functions being among the most popular \citep{feldmann2023flexible}.

\section{Smoothness selection for nonparametric Markov-switching models}

\label{sec:smoothness}

\subsection{Penalized likelihood for a given penalty strength}
\label{subsec:PML}

All models formulated above can be fitted to data by direct numerical maximum likelihood estimation. More specifically we employ the so-called forward algorithm to calculate the likelihood recursively \citep{zucchini2016hidden}, leading to the closed-form expression
\begin{equation}
    \label{eqn:likelihood}
    \mathcal{L}(\bm{\theta}) = \bm{\delta}^{(1)} \bm{P}(x_1) \bm{\Gamma}^{(2)} \bm{P}(x_2) \bm{\Gamma}^{(3)} \dotsc \bm{\Gamma}^{(T)} \bm{P}(x_T) \bm{1},
\end{equation}
with initial distribution $\bm{\delta}^{(1)}$, transition probability matrix $\bm{\Gamma}^{(t)}$, the matrix of state-dependent densities or probability mass functions $\bm{P}(x_t) = \text{diag}\bigl(f_1(x_t), \dotsc, f_N(x_t)\bigr)$, and $\bm{1}$ a column vector of ones. The dependence of $\bm{\Gamma}^{(t)}$ and $\bm{P}(x_t)$ on the parameters to be estimated is determined by the specific model formulation from Section \ref{sec:model formulation}. To avoid numerical underflow we use a scaled version of the forward algorithm, thus calculating the log-likelihood $l(\bm{\theta})$. For practical implementation, we use the R package \texttt{LaMa} \citep{koslikLaMa2024} mainly for its convenience and computational efficiency. Furthermore, all constrained parameters are transformed via an invertible and differentiable link function to allow for unconstrained maximization.
%each model presented in Section \ref{sec:model formulation} can be estimated by numerically maximizing $l(\bm{\theta})$ using any suitable maximization routine like R's \texttt{nlm} \citep{schnabel1985modular}.
We then summarise the unconstrained versions of all unpenalized 
%non-spline 
parameters not associated with splines in a vector $\bm{a} \in \mathbb{R}^d$ and all non-zero spline coefficients in a vector $\bm{b} = (\bm{b}_1, \dotsc, \bm{b}_p) \in \mathbb{R}^{pK}$ --- now using a single index to ease notation --- such that the full parameter vector is $\bm{\theta} = (\bm{a}, \bm{b})$. The number of smooths $p$ depends on the specific model class, e.g.\ $p = N$ for (univariate) density estimation, $p = N Q$ for MS-GAM, and $p = N(N-1) Q$ for nonparametric modeling of the transition probabilities. 

% As alluded to earlier, when incorporating splines into the model, it is necessary to control model flexibility to avoid overfitting. 
Instead of explicitly selecting the number of basis functions, to control model flexibility, we fix a sufficiently large $K$ and control 
%the model flexibility, i.e.\ 
the wiggliness of the estimated functions by penalizing the curvature of the nonparametric state-dependent densities $f_i$ (for models as in Section \ref{subsec:nonparametric_densities}) or the smooth functions $s_i$ (for models as in Sections \ref{subsec:MSGAM} and \ref{subsec:transprobs}). Different penalties for controlling the variability can be represented as a quadratic form, leading to the penalized log-likelihood
% Due to the representation of each smooth term as a linear combination, for a well-behaved basis, a penalty on the integrated squared second derivative
% $\int s''(x)^2 \,dx$
% can be represented by a quadratic form \citep{wood2017generalized}, leading to the penalized log-likelihood
\begin{equation}
    \label{eq:penalizedllk}
    l_p(\bm{\theta}; \bm{\lambda}) = l(\bm{\theta}) -
\frac{1}{2} \sum_{i=1}^p \lambda_i \bm{b}_i^\intercal \bm{S}_i \bm{b}_i,
\end{equation}
where $\bm{S}_i$ is a penalty matrix of known coefficients %depending on the basis 
and $\bm{\lambda} = (\lambda_1, \dotsc, \lambda_p)$ is a vector of smoothing strengths determining the amount of penalization for each smooth. 
For example, we can directly penalize the integrated squared second derivative $\int s''(x)^2 \,dx$ which, for a well-behaved basis, can be represented by a quadratic form \citep{wood2017generalized}. In this case, $\bm{S_i}$ has entries $\int B_j^{\prime \prime}(x) B_k^{\prime \prime}(x) \;dx$. Alternatively, we can
%As an example, 
consider penalizing the squared second-order differences of the spline coefficients, proposed by \citet{eilers1996flexible}: For this, let $\Tilde{\bm{D}_2}$ denote the second-order difference matrix of dimension $(K-1) \times (K+1)$. When setting $b^{(i)}_0 = 0$ for identifiability, we can omit the linear combination with $b^{(i)}_0$ and set $\bm{D}_2$ =  $\Tilde{\bm{D}}_2 [,\{ 1, \dotsc, K\}]$ by dropping the first column corresponding to $b^{(i)}_0$. Thus, the second-order differenced coefficients are $\bm{D}_2 \bm{b}_i$ and summing their squares amounts to the quadratic form $\bm{b}_i^\intercal \bm{D}_2^\intercal \bm{D}_2 \bm{b}_i$, hence $\bm{S}_i = \bm{D}_2^{\intercal} \bm{D}_2$. 

The procedure detailed in the following section can be applied for any penalized spline formulation as long as the penalty can be written in a quadratic form. For practical implementation, we will mostly use the design and penalty matrices provided by the highly optimized R package \texttt{mgcv} \citep{wood2015package}, as these already account for identifiability.

For a given penalty strength $\bm{\lambda}$, $l_p(\bm{\theta}; \bm{\lambda})$
%Equation \eqref{eq:penalizedllk} 
can then be maximized numerically using any suitable maximization routine like R's \texttt{nlminb} \citep{gay1990usage} or \texttt{optim} \citep{rstats}. 
To increase performance and accuracy, we make the penalized likelihood function compatible with automatic differentiation using the novel R package \texttt{RTMB} \citep{kristensen2024rtmb}.

\subsection{Existing work on selecting the penalty strength}

\label{subsec:existing}

While the previous section detailed model estimation \textit{given} penalty strengths $\bm{\lambda} = (\lambda_1, \dotsc, \lambda_p)$, in practice, the more challenging task lies in selecting appropriate penalty strengths in a data-driven way. To date, smoothness selection for nonparametric HMMs in underdeveloped \citep{langrock2017markov, langrock2018spline} and existing approaches typically involve a grid search over possible penalty strengths employing cross-validation or information criteria to compute the score. However, cross-validation for time-series data is not straightforward \citep[][]{celeux2008selecting} and information criteria have been shown to be unreliable for model selection in HMMs \citep{pohle2017selecting}.

Most importantly, however, grid-search approaches suffer from the curse of dimensionality when multiple smooths are included in the model. % which is very typical in HMM analyses. 
Suppose for example that a 3-state HMM is to be fitted, with a single covariate nonparametrically affecting the t.p.m.\ -- with a grid of only five values for each of the six smooths, a 10-fold CV would theoretically require $5^6 \cdot 10 = 156250$ model fits. 
This can be remedied by using greedy search methods, for example, starting at some grid value, calculating the score for all direct neighbors, moving to the best one, and repeating \citep{langrock2015nonparametric}. However, such a greedy approach increases the chances of missing the optimal penalty strength. 
% For example, including three smooths and searching a grid of only five values via a 10-fold CV requires $5^3 \cdot 10 = 1250$ model fits. %, while a grid of five values can be considered relatively coarse, making suboptimal solutions probable. 
% We want to emphasize that a number of three smooths is relatively small for an HMM: For example, linking the t.p.m.\ to one covariate only via smooth functions, already results in 6 penalty strength parameters. 
Overall, these considerable challenges associated with smoothness selection severely hamper the practical use of spline-based nonparametric HMMs.

More recently, \citet{michelot2022hmmtmb} proposed an alternative to the above approaches by treating the spline coefficients as random effects and the penalty as their multivariate normal distribution.
%This perspective has been around for a long time for GAMs \citep{}. 
Smoothness selection for penalized splines can then be conducted via marginal maximum likelihood estimation, integrating out the random effects, which is asymptotically justified \citep{kauermann2009some}. The marginal likelihood of the data $\bm{x}$, as a function of fixed effects $\bm{a}$ and the smoothing parameter $\bm{\lambda}$ then has the general form
\begin{equation}
\label{eq:marginal_lik}
    \mathcal{L}(\bm{a}, \bm{\lambda}) = f_{\bm{a}, \bm{\lambda}}(\bm{x}) = \int f_{\bm{a}}(\bm{x} \mid \bm{b}) f_{\lambda}(\bm{b}) \;d\bm{b},
\end{equation}
where $f_a(\bm{x} \mid \bm{b})$ is the likelihood as a function of $\bm{a}$ and $\bm{b}$. $f_{\lambda}(\bm{b})$ is a suitable multivariate normal distribution corresponding to the penalty in \eqref{eq:penalizedllk} which we will discuss in more detail in Section \ref{subsec:qREML}.
% It is a well-known fact, that marginal maximum likelihood estimation typically produces smoother function estimates than CV approaches \citep{wood2011fast} which is indeed desirable for HMM applications due to the high model complexity with latent components typically leading to less-stable models, compared to e.g.\ GAMs.
Marginal maximum likelihood estimation can be conducted using the R package \texttt{TMB} \citep{kristensen2016tmb} to evaluate the marginal log-likelihood (and its gradient) for a very general class of random effects, including the special case of spline smoothing. 
The main building block of \texttt{TMB} for this is the so-called \textit{Laplace approximation} of the high-dimensional integral, based on a second-order Taylor expansion around the mode $\Hat{\bm{b}}(\bm{a}, \bm{\lambda})$ of the joint likelihood, i.e.\ the integrand in \eqref{eq:marginal_lik}, with respect to $\bm{b}$. As this mode depends on both the fixed effects $\bm{a}$ as well as the penalty strength $\bm{\lambda}$, optimizing \eqref{eq:marginal_lik} numerically using a Newton-Raphson algorithm requires an inner optimization over $\bm{b}$ for each outer iteration. Thus, the Laplace approximation leads to a nested optimization problem and optimization is only feasible because the key development of \texttt{TMB} is the use of automatic differentiation (AD) \citep{fournier2012ad}, providing fast and accurate gradient information.
Nonetheless, this method remains computationally extremely demanding while sometimes being numerically unstable. 
% , for which in each Newton-Raphson step in the optimization over the fixed parameters $\bm{a}$ and the penalty strength $\bm{\lambda}$, the joint likelihood of $\bm{a}$, $\bm{\lambda}$ and $\bm{b}$ needs to be maximized with respect to $\bm{b}$, 
% rendering the approach computationally extremely demanding. The computational burden indeed prohibits practical usability if derivatives are obtained via finite differencing and the key development of \texttt{TMB} is the use of automatic differentiation (AD) \citep{fournier2012ad} within the nested optimization, rendering the approach feasible. 
%The disadvantage is that the user needs to define the joint log-likelihood function in C++ which may often be inconvenient for practitioners used to writing tailored R code to optimize numerically. 
Additionally, not only the estimation of the penalty strength $\bm{\lambda}$, but also of the fixed effects $\bm{a}$, depends on the quality of the Laplace approximation --- which, in general, is difficult to assess.
%(we will try to give some intuition in Section \ref{sec:simulation}).

%, which is difficult to assess and may not be good for complicated HMM likelihoods \citep{HMM likelihoods not well behaved}.

However, for the special case of spline smoothing, the intense computational overhead required for the nested optimization is not necessary when employing two key changes to the above procedure which will be detailed in Section \ref{subsec:qREML}.
First, instead of integrating out $\bm{b}$ only, it is convenient to treat $\bm{a}$ also as a random effect and integrate it out, which is known as restricted maximum likelihood (REML). Second, the now simplified outer optimization problem can be approximately solved by closed-form updates to $\bm{\lambda}$ when treating the modes $\Hat{\bm{a}}$ and $\Hat{\bm{b}}$ as constants not depending on $\bm{\lambda}$ in each outer iteration. %and only partially differentiate terms that explicitly depend on $\bm{\lambda}$.
The algorithm we suggest can then be carried out regardless of derivatives being obtained via finite differencing or AD with the latter of course being faster and more reliable. The procedure produces adequately smooth function estimates while being computationally much less demanding, very robust, and extremely easy to implement for anyone familiar with direct numerical maximum likelihood estimation for Markov-switching models.

\subsection{qREML for Markov-switching models}

\label{subsec:qREML}

As alluded to above, we also build on the random effects framework, following \citet{wahba1985comparison, schall1991estimation, fahrmeir2004penalized, wood2011fast, schellhase2012density}. The spline coefficients are treated as random effects with a potentially improper Gaussian distribution:
$$\bm{b}_i \sim \mathcal{N}(0, \lambda_i^{-1} \bm{S}_i^-),$$
where $\bm{S}_i^-$ is the generalized inverse of $\bm{S}_i$, because $\bm{S}_i$ might not be of full rank, making the above normal distribution improper. For example, when penalizing squared second-order differences of the coefficients, linear trends are not penalized, hence lie in the null space of $\bm{S}_i$. We seek to estimate the model by integrating out the random effects. To overcome the issue of nested numerical optimization we need to simplify the outer optimization problem. Thus, we employ a mathematical trick by not only treating the spline parameters $\bm{b}$ as random effects but also regarding the remaining model parameters $\bm{a}$ as such, an idea already proposed by \citet{laird1982random} for mixed models, known as \textit{restricted maximum likelihood} (REML). For $\bm{a}$ we assume what from a Bayesian perspective would be called a \textit{flat prior} or the limit of a multivariate normal distribution with \textit{infinite} variance, so with precision practically equal to zero, thus not penalizing the fixed effects. In the limit, $\bm{a}$ does not have a density, but for the remaining calculations, we could also assume a finite, but very large variance that is not estimated. Moreover, we assume that $\bm{a}$ is independent of $\bm{b}$, and clearly as implied by the additivity of the penalty, the components of $\bm{b}$ are mutually independent. 

To address the potential rank-deficiency of $\bm{S}_i$, we decompose it into
$$
\bm{S}_i = \bm{U}_i \bm{\Lambda}_i \bm{U}_i^\intercal,
$$
where $\bm{U}_i$ is an orthonormal matrix and 
$$\bm{\Lambda}_i = \begin{pmatrix}
    \bm{\Lambda}_i^* & \bm{0} \\
    \bm{0} & \bm{0}
\end{pmatrix},$$
where $\bm{\Lambda}_i^*$ contains the $K - m_i$ positive eigenvalues of $\bm{S}_i$. Separating the columns of $\bm{U}_i$ into $\bm{U}_i = (\bm{U}_i^*, \bm{U}_i^0)$, the columns of $\bm{U}_i^*$ matching $\bm{\Lambda}_i^*$, we now reparametrize in terms of $\bm{b}_i^* = \bm{U}_i^{* \intercal} \bm{b}_i$ and $\bm{b}_i^0 = \bm{U}_i^{0 \intercal} \bm{b_i}$, with the suggestive notation that $\bm{b}_i^*$ is the part of the coefficient vector that is penalized while $\bm{b}_i^0$ lies in the penalty nullspace. If the eigenvalues associated with $\bm{b}_i^0$ were not zero but slightly positive --- i.e.\ there would be a small penalty on the nullspace --- $\bm{b}_i^0$ would also admit a normal distribution with a large variance, hence we treat it exactly as $\bm{a}$. Moreover, from the block-diagonal structure of $\bm{\Lambda}_i$ we see that $\bm{b}_i^*$ and $\bm{b}_i^0$ are a priori independent.

Following \citet{laird1982random}, we now build the integral of the joint likelihood but in contrast to equation \eqref{eq:marginal_lik} integrate out $\bm{a}$, $\bm{b}^0 = (\bm{b}_1^0, \dotsc, \bm{b}_p^0)$ and $\bm{b}^* = (\bm{b}_1^*, \dotsc, \bm{b}_p^*)$ to arrive at the marginal likelihood of $\bm{\lambda}$ only,
$$
\mathcal{L}(\bm{\lambda}) = \Bigl( \prod_{i=1}^p \det(\lambda_i \bm{\Lambda}_i^*)^{1/2} \Bigr) \cdot C \cdot \int \exp \bigl(l(\bm{a}, \bm{b}) \bigr) \exp \Bigl(-\frac{1}{2} \sum_{i=1}^p \lambda_i \bm{b}_i^\intercal \bm{S}_i \bm{b}_i \Bigr) \,d\bm{a}\, d\bm{b}^0\, d\bm{b}^*.
$$
As this integral is numerically intractable, we perform a Laplace approximation around the mode $(\Hat{\bm{a}}, \Hat{\bm{b}})$ which --- after dropping additive constants --- yields the approximate marginal log-likelihood
\begin{equation}
    \label{eq:marginalLLK}
l(\bm{\lambda}) \approx \frac{1}{2} \sum_{i=1}^p (K - m_i) \log(\lambda_i) + l(\Hat{\bm{a}}, \Hat{\bm{b}}) - \frac{1}{2} \sum_{i=1}^p \lambda_i \Hat{\bm{b}}_i^\intercal \bm{S}_i \Hat{\bm{b}}_i - \frac{1}{2} \log \det (\bm{V}^\intercal \bm{J}_p(\bm{\lambda}) \bm{V}), 
\end{equation}
where $\bm{J}_p(\bm{\lambda}) = -\nabla^2 l_p(\bm{a}, \bm{b}; \bm{\lambda}) \rvert_{\Hat{\bm{a}}, \Hat{\bm{b}}}$ and $\bm{V} = \text{bdiag}(\bm{I}_d, \bm{U}_1, \dotsc, \bm{U}_p)$. 
From this point, we proceed analogously to \citet{schellhase2012density}, whose ideas originated from \citet{schall1991estimation}. We \textit{partially} differentiate \eqref{eq:marginalLLK} with respect to $\lambda_i$, treating $\Hat{\bm{a}}$ and $\Hat{\bm{b}}$ as fixed constants, yielding
\begin{align*}
   \pdv{l(\bm{\lambda})}{\lambda_i} &\approx \frac{K - m_i}{2 \lambda_i} - \frac{1}{2} \Hat{\bm{b}}_i^\intercal \bm{S}_i \Hat{\bm{b}} - \frac{1}{2} \Tr \bigl\{ \bigl( (\bm{V}^\intercal \bm{J}_p(\bm{\lambda}) \bm{V})^{-1} \bigr)_{ii} \bm{U}_i^\intercal \bm{S}_i \bm{U}_i \bigr\}\\
   &= \frac{K - m_i}{2 \lambda_i} - \frac{1}{2} \Hat{\bm{b}}_i^\intercal \bm{S}_i \Hat{\bm{b}} - \frac{1}{2} \Tr \bigl\{ \bm{U}_i (\bm{J}_p(\bm{\lambda})^{-1})_{ii} \bm{U}_i^\intercal \bm{U}_i^\intercal \bm{S}_i \bm{U}_i \bigr\}
\end{align*}
where $()_{ii}$ refers to the indices of $\bm{b}_i$. For practical implementation, we approximate the trace expression by $\Tr \bigl( (\bm{J}_p(\bm{\lambda})^{-1})_{ii} \bm{S}_i\bigr)$, yielding
%By approximating $\bigl( (\bm{V}^\intercal \bm{J}_p(\bm{\lambda}) \bm{V})^{-1} \bigr)_{ii}$ by $\bm{U}_i^\intercal (\bm{J}_p(\bm{\lambda})^{-1})_{ii} \bm{U}_i$ we get
$$
\pdv{l(\bm{\lambda})}{\lambda_i} \approx - \frac{1}{2} \Hat{\bm{b}}_i^\intercal \bm{S}_i \Hat{\bm{b}} - \frac{1}{2 \lambda_i} \Bigl( K - m_i - \lambda_i \Tr \bigl( (\bm{J}_p(\bm{\lambda})^{-1})_{ii} \bm{S}_i \bigr) \Bigr).
$$
From this, we can construct an estimating equation for $\lambda_i$ as
\begin{equation}
\label{eq: updating_equation}
    \lambda_i = \frac{K - \lambda_i \Tr\bigl( (\bm{J}_p(\bm{\lambda})^{-1})_{ii} \bm{S}_i \bigr) - m_i}
    %\frac{\Tr \bigl( \bm{I}_K - \lambda_i (\bm{J}_p(\bm{\lambda})^{-1})_{ii} \bm{S}_i \bigr) - m_i}
    {\Hat{\bm{b}}_i^\intercal \bm{S}_i \Hat{\bm{b}}}.
\end{equation}
As both sides depend on $\bm{\lambda}$, this suggests an iterative procedure, calculating the mode 
%$(\Hat{\bm{a}}, \Hat{\bm{b}})$ 
of $l_p(\bm{\theta}; \bm{\lambda})$, with $\bm{\theta} = (\bm{a}, \bm{b})$, given the current penalty strength $\bm{\lambda}$ and then updating the penalty strength based on Equation \eqref{eq: updating_equation}. Calculating the mode corresponds to fitting the model via penalized maximum likelihood estimation as explained in Section \ref{subsec:PML}. As we only partially differentiate \eqref{eq:marginalLLK}, instead of acknowledging the dependence of the penalized MLE on $\bm{\lambda}$ and obtaining derivatives of $\Hat{\bm{a}}$ and $\Hat{\bm{b}}$ w.r.t.\ to $\bm{\lambda}$ (similar to \cite{wood2011fast}), we obtain an approximation of the full REML procedure, hence referring to it as \textit{quasi restricted maximum likelihood (qREML)}.

We can gain some intuition for Equation \eqref{eq: updating_equation} by rewriting the numerator as
$$
\Tr \bigl( \bm{I}_K - \lambda_i (\bm{J}_p(\bm{\lambda})^{-1})_{ii} \bm{S}_i \bigr) - m_i
$$
% \begin{equation}
%     \label{calculation}
%     \pdv{l(\bm{\lambda})}{\lambda_i} \approx - \frac{1}{2} \Hat{\bm{b}}_i^\intercal \bm{S}_i \Hat{\bm{b}} - \frac{1}{2 \lambda_i} \Bigl( \Tr \bigl( \bm{I}_K - \lambda_i (\bm{J}_p(\bm{\lambda})^{-1})_{ii} \bm{S}_i \bigr) - m_i \Bigr).
% \end{equation}
and replacing $\bm{I}_K$ by $(\bm{J}_p(\bm{\lambda})^{-1})_{ii} \bm{J}_p(\bm{\lambda})_{ii}$ --- which holds approximately if the off-diagonal entries are small. The trace expression then becomes
$$
\Tr \bigl( (\bm{J}_p(\bm{\lambda})^{-1})_{ii} (\bm{J}_p(\bm{\lambda})_{ii} - \lambda_i \bm{S_i})\bigr) = \Tr \bigl( (\bm{J}_p(\bm{\lambda})^{-1})_{ii} \bm{J}(\bm{0})_{ii} \bigr),
$$
with $\bm{J}(\bm{0})$ denoting the negative Hessian of the unpenalized likelihood at $\Hat{\bm{a}}$ and $\Hat{\bm{b}}$. The equality is true because of the additive structure of the penalized log-likelihood w.r.t.\ the penalty. Without the index restriction, the expression on the right-hand side is known as the \textit{effective degrees of freedom}, defined as the trace of the product of the inverse Fisher information matrix for the penalized likelihood and the Fisher information matrix for the unpenalized likelihood \citep{gray1992flexible}. Hence, we can interpret the trace expression as the approximate degrees of freedom for smooth $i$. Subtracting $m_i$ corrects for the penalty matrix potentially being rank-deficient.

The computational savings stem from the fact that jointly optimizing the penalized likelihood with respect to $\bm{a}$ and $\bm{b}$ is not considerably more demanding than optimizing with respect to $\bm{b}$ for fixed $\bm{a}$, as the number of fixed effects is typically small compared to the number of random effects. However, crucially, the outer optimization problem becomes substantially easier mathematically when we only optimize the marginal likelihood with respect to $\bm{\lambda}$ compared to jointly optimizing it for $\bm{\lambda}$ and $\bm{a}$, because in the special case of penalized spline smoothing --- as compared to general random effects --- the penalty is linear in the penalty strength parameters. As it is necessary to compute the effective degrees of freedom for each smooth anyway for Equation \ref{eq: updating_equation}, obtaining the model's (exact) total degrees of freedom from the last iteration to compute information criteria like AIC and BIC for model comparisons is also straightforward.
% true because of trace and block diagonal structure of big penalty matrix! tr(I_pK - J_p^-1 bigS)
Equation \eqref{eq: updating_equation} also reveals the interesting relationship that, at convergence, the optimal penalty strength effectively balances the penalty added to the log-likelihood for smooth $i$ with the effective number of parameters attributed to this smooth.

\subsection{Comparison with existing approaches and some remarks}

The selection of the smoothness penalty via cross-validation or information criteria will typically only be feasible using parallelisation with respect to the different penalty strengths being tested, but will in any case be very computer-intensive. In contrast, in qREML the smoothness penalty selection is part of the (single) model fitting exercise. In addition, the iterative nature of the qREML approach can be exploited by initialising the parameter vector $\bm{\theta} = (\bm{a}, \bm{b})$ with its estimate from the previous outer iteration for each inner optimization.
% While compared to cross-validation and information criteria, qREML holds the obvious disadvantage of not being parallelizable due to its iterative nature, the latter can be exploited: For each inner optimization, we can initialize the parameter vector 
% $\bm{\theta} = (\bm{a}, \bm{b})$ with its estimate from the previous outer iteration. 
This has two important advantages. First, it drastically improves estimation speed as the number of iterations necessary for convergence can drop substantially due to the new solution being close to the old one. Second, globally initializing with a fairly large penalty strength, the algorithm provides a smooth transition between inflexible, hence very stable models and the resulting optimal fit, potentially reducing the risk of convergence to local optima (see Figures \ref{fig:energy_oil_modseq} and \ref{fig:elephant_model_sequence} in the Appendix for a visual illustration).

We want to highlight that treating the fixed parameters as random effects (with infinite variance) has the advantage that their estimation, for a given penalty strength, is exact and does not depend on the quality of the Laplace approximation which is generally difficult to access. The approximation quality influences only the optimal penalty choice, which is considerably less important for the final model fit and inference than the fixed effect estimates. In the case studies, we will demonstrate that once the penalty strength is relatively close to its optimum, there are almost no visible changes in the estimated smooth functions anymore.

In the context of generalized additive models (GAMs) and generalized linear mixed models (GLMMs), the use of cross-validation, marginal maximum likelihood, and restricted maximum likelihood as methods for nonparametric smoothing and mixed modeling has been thoroughly explored. For instance, \citet{reiss2009smoothing} compared REML with generalized cross-validation (GCV), demonstrating that REML typically imposes a stricter penalty on overfitting than GCV.

\citet{wood2011fast} states that single-iteration methods, like the one proposed here, generally do not ensure convergence to a fixed $\Hat{\bm{\lambda}}$, but also highlights that full REML and full marginal ML approaches are substantially more computationally intensive than single-iteration methods. We contend that, for spline-based nonparametrics, achieving highly precise estimates of $\Hat{\bm{\lambda}}$ may be less critical than the practical speed improvements offered by qREML. Moreover, in the data scenarios we examined, achieving convergence with a relatively high tolerance level did not pose any significant issues, though this may warrant further exploration in future studies.

% Lastly, for some historical context,  compare marginal ML to generalized cross-validation (GCV) and indicate that both options may converge to local optima, but GCV may be less stable. For HMMs, GCV does not provide a viable alternative anyway, and fitting HMMs is always prone to numerical difficulties, hence making the convergence considerations much less important than for GAMs. Moreover, for HMMs, the computational advantage is severely more important as model fitting via numerical maximization is costly and a single model fit can be time-consuming, depending on the model complexity and the data situation at hand. Thus, especially when several smooths should be included in the model, marginal ML reduces the computational burden substantially, compared to grid searches on a high-dimensional grid, necessary for CV. Nonetheless, we want to point out that non-convergence is an issue to be considered, and trace plots of the penalty strengths should always be inspected for evidence of convergence.

\subsection{Practical implementation}
\label{subsec:practical}

For practical implementation, we include the two functions \texttt{qreml()} and \texttt{penalty()} in the R package \texttt{LaMa} \citep{koslikLaMa2024}. 
These build on the R package \texttt{RTMB} \citep{kristensen2024rtmb}, an interface to \texttt{TMB} from R, to allow for automatic differentiation in the penalized likelihood estimation while not using the package's Laplace approximation functionalities.
%that allows for automatic differentiation while the user only needs to specify the likelihood function in plain R code. 
To use \texttt{qreml()}, the user merely needs to implement the \textit{negative penalized log-likelihood function} that is compatible with \texttt{RTMB} and use the \texttt{penalty()} function to compute all quadratic form penalties. The penalized likelihood function can then be passed to \texttt{qreml()} and after specifying which parameters are spline coefficients, the qREML algorithm is used to find the optimal penalty strength parameter. The outer optimization is terminated once the largest absolute value of relative change in $\bm{\lambda}$ falls below a threshold of $10^{-4}$ and the \texttt{qreml()} function is designed to provide a very similar workflow to the estimation of random effects models with \texttt{RTMB} that uses the full Laplace approximation. The inner optimization is done by \texttt{optim()} \citep{rstats} with the BFGS method because it showed the overall best performance regarding both speed and stability. For each $\bm{\lambda}$ update, the Hessian is obtained from \texttt{optim()} via finite differencing of the AD gradient because this method was much faster than computing the Hessian using AD with \texttt{RTMB} after convergence.

For uncertainty quantification, by default, the algorithm returns the negative Hessian of the penalized log-likelihood at convergence which can then be used to quantify uncertainty in $(\Hat{\bm{a}}, \Hat{\bm{b}})$ based on maximum likelihood theory \textit{conditional} on the estimated penalty strength $\Hat{\bm{\lambda}}$. 
However, after optimization, \texttt{qreml()} also automatically constructs and returns the full marginal log-likelihood object with \texttt{RTMB} --- adding normalization constants for the Gaussian distributions of the spline coefficients missing from the penalized likelihood. \texttt{RTMB}'s \texttt{sdreport()} can then calculate the joint Hessian of all parameters and the penalty strength, hence allowing for joint uncertainty quantification in $(\Hat{\bm{a}}, \Hat{\bm{b}}, \Hat{\bm{\lambda}})$ \citep[see][Section 6.10]{wood2017generalized}%, thereby allowing for \textit{joint} uncertainty quantification using the observed joint Fisher information.

\subsection{Simulation experiments}
To demonstrate the general practicality of the qREML approach and to investigate the number of iterations required for convergence and its dependence on sample size, we conducted simulation experiments, the results of which are reported in Appendix \ref{sec:simulation}. In these experiments, the smoothness penalty was chosen adequately in almost all simulation runs, reliably producing satisfactory function estimates. Notably, the convergence speed of the qREML algorithm increases with increasing sample size $T$. This can be explained by the likelihood function increasingly resembling the shape of a multivariate Gaussian as $T$ grows, which in turn improves the accuracy of the Laplace approximation.

\section{Case studies}

The following three case studies demonstrate the practical use of the proposed qREML approach for the different types of nonparametric Markov-switching models as presented in Section \ref{sec:model formulation}. All code and data for full reproducibility of the analyses is available at \url{https://github.com/janoleko/qREML}.

\label{sec:case studies}

% \subsection{Whales: depth-displacement per dive}

\subsection{Caracaras activity levels}

Our first case study focuses on the activity levels of juvenile striated caracaras (Phalcoboenus australis) on Saunders Island in the Falkland Islands, 
%home to an estimated 50 to 100 caracaras year-round, 
as documented by \citet{harrington2020seasonal}. From 2017 to 2019, researchers deployed animal-borne accelerometers and GPS data loggers on 27 juvenile caracaras during both summer and winter seasons, in an attempt to investigate their behavioral response to the presence and absence of seasonally migratory prey. The detailed methodology can be found in \citet{harrington2020seasonal}. 
% For such analyses, it is thus critical to employ a model able to accurately characterize the behavioral states from the noisy observed data. 
For simplicity, in our analysis, we focus on only one individual with deployment ID \texttt{20190221\_TW9\_M57}, downsampling its acceleration data to 0.1 Hz, and hence $T = 8640$ observations in total, as the extremely high temporal resolution is not actually needed for reliable inference on the animal's behavior. We use the vector dynamic body acceleration (VeDBA) as considered also by \citet{harrington2020seasonal}, and convert it to the log scale to avoid complications with the otherwise restricted support. The original paper identified four distinct behavioral states, with states 1 and 2 both representing resting behavior, where state 2 corresponded to resting with noise. 

\begin{figure}
    \centering
    \includegraphics[width=1
\textwidth]{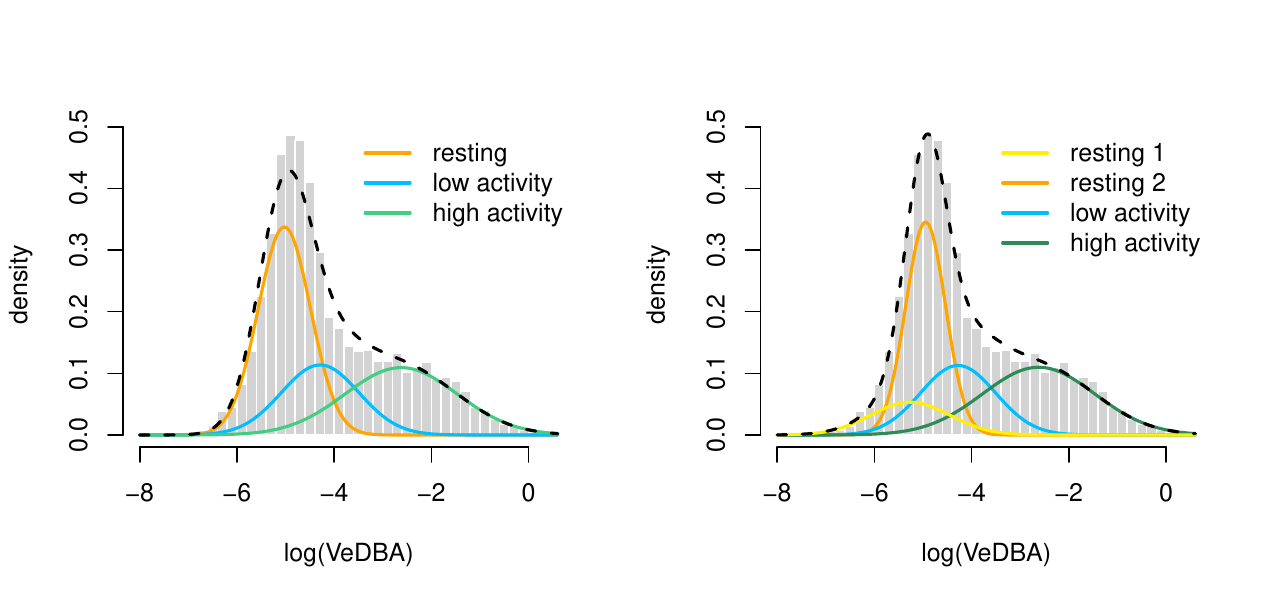}
    \caption{Histogram of the acceleration data complemented with the state-dependent distributions and the marginal distribution of the 3-state (left panel) and 4-state (right panel) normal HMM fitted to the caracara data.}
    \label{fig:caracaras_simple}
\end{figure}

To establish a baseline, we fitted 3-state and 4-state HMMs with normal emission distributions, finding a similar interpretation to \citet{harrington2020seasonal}, with states corresponding to resting behavior, low activity, and high activity. The state-dependent distributions of the estimated models are depicted in Figure \ref{fig:caracaras_simple}. Our results indicate that the 4-state model, which includes two distinct resting states, fits the data considerably better in particular capturing the marginal distribution much more precisely than the 3-state model. The Akaike Information Criterion (AIC) and the Bayesian Information Criterion (BIC) favor the 4-state model, with scores of 21791.7 and 21933.0, respectively, compared to 22044.1 and 22128.9 for the 3-state model.

While \citet{harrington2020seasonal} label their first two states as ``resting'' and ``resting with noise'', this distinction is less clear for the subset of the data considered here. The state labeled ``resting 1'' includes both smaller and larger VeDBA values compared to the state labeled ``resting 2''. Consequently, there is no compelling mechanistic reason to separate these two states apart from achieving a better model fit. The high overlap in the state-dependent distributions complicates the interpretation of the 4-state model, as the emission distribution of the first state encompasses both very small acceleration values and values that could also belong to the second resting state. Therefore, the analysis might benefit from a more flexible model that can accurately represent the true state-dependent distributions while potentially providing better state separation.

As an alternative, we thus fit a 3-state nonparametric HMM to the same data, modeling the nonparametric state-dependent distributions as a finite combination of standardized cubic B-spline basis functions with non-negative weights summing to one, placing the knots equidistantly in the range of the log(VeDBA). For each state-dependent density, we use 25 B-spline basis functions to enable sufficient flexibility, penalizing the squared second-order differences of the spline coefficients. To fit the model, we used the qREML algorithm developed in Section \ref{sec:smoothness}, initializing the penalty strength as $\bm{\lambda}_0 =(30,30,30)$.
% Unlike the regression-type estimation of smooth functions, for 
For density estimation we cannot use an excessive penalty strength in the first iteration, as the model needs to be able to represent distinct state-dependent distributions, which would be hindered by an overly strong penalty. We leave the convergence tolerance at its default, terminating the algorithm once the largest absolute value of relative change in the penalty strength falls below $10^{-4}$. Convergence was achieved in eight iterations and model fitting took 29.93 seconds on an Apple M2 chip. Fitting the model with the full REML procedure (which can be done natively with \texttt{RTMB}) took 176.08 seconds and model fitting using marginal ML took 390.73 seconds.
With qREML, the penalty strength parameter was estimated to be $\Hat{\bm{\lambda}} = (2.24, 8.49, 12.43)$.

\begin{figure}
    \centering
    \includegraphics[width=1
\textwidth]{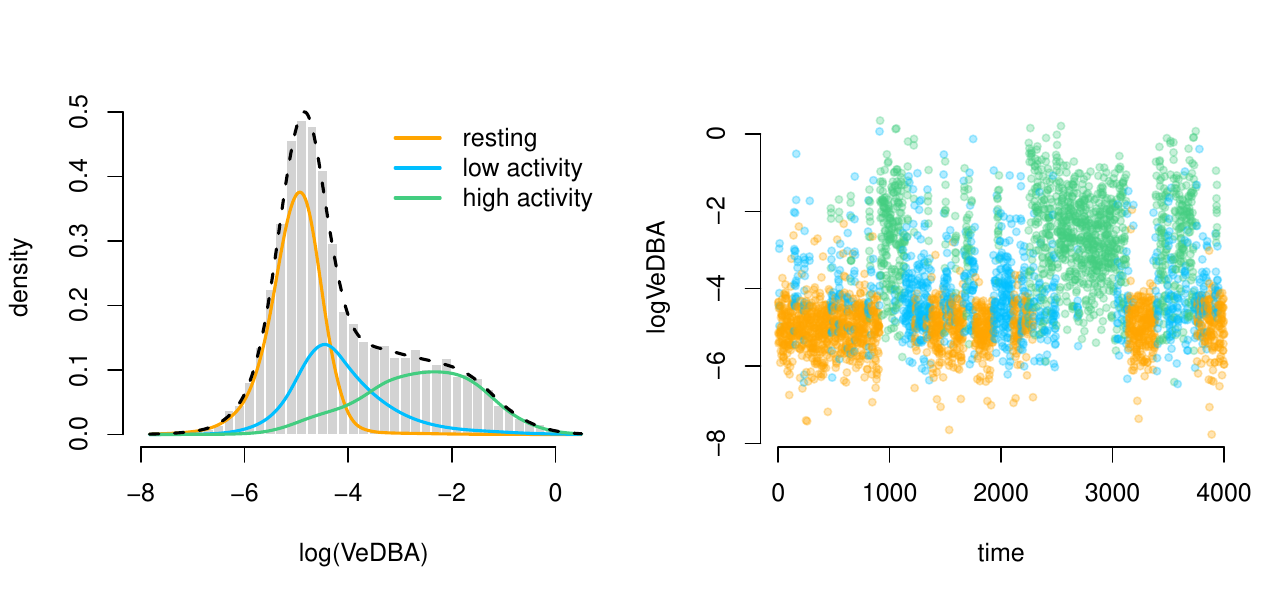}
    \caption{Histogram of the acceleration data complemented with the state-dependent distributions and the marginal distribution of the 3-state nonparametric HMM fitted to the caracara data (left panel) and part of the time series of vector dynamic body acceleration colored according to the Viterbi-decoded state sequence (right panel)}
    \label{fig:caracaras_spline}
\end{figure}

Figure \ref{fig:caracaras_spline} shows the estimated state-dependent distributions of the HMM with nonparametric emission distributions. We see that the marginal distribution fits the empirical distribution exceptionally well, achieving this fit with only three states. AIC, with a value of 21776.29, also favors the nonparametric model over the previous parametric models while the BIC has a value of 22010.48, actually preferring the four-state model. Conditional AIC and BIC are calculated from the unpenalized log-likelihood at the penalized MLE, with effective parameter counts derived as explained in Section \ref{subsec:qREML}.
The state-dependent distribution of the resting state is slightly left-skewed, which helps to capture the left tail of the empirical distribution, rendering a second resting state unnecessary. 
The state-dependent distributions in both active states are also estimated to be asymmetric.
%The estimated state-dependent distribution of the low activity state is asymmetric and exhibits a high kurtosis enabling better state separation and clearer interpretation of the states.

This finding supports the conclusions of \citet{pohle2017selecting} who found that inadequately specified state-dependent distributions can lead to an overestimation of the number of states. Such overestimation can result in less interpretable models and invalidate statistical inference.

% \subsection{nonparametric density estimation for tuco tuco acceleration}

% \begin{figure}[h]
%     \centering
%     \includegraphics[width=0.7\textwidth]{figs/tuco_marginal.pdf}
%     \caption{Caption}
%     \label{fig:enter-label}
% \end{figure}

% \begin{itemize}
%     \item triaxial accelerometer data, collected from four wild tuco-tucos, captured near Anillaco, La Rioja, Argentina. \citep{}
%     \item here, we only consider animal 2
%     \item data collected over three days in March 2019, sampled at a rate of 10Hz with a sensitivity of 4g.
%     \item Data from the days of capture and release was discarded to avoid any side effects caused by disturbing the animals in their natural habitat
%     \item subsampled by a factor of 30
%     \item raw acceleration data hard to analyse, we computed vector dynamic body acceleration (VeDBA) --- which is the Euclidean norm of the three measurements.
%     \item Furthermore, converted to log scale to better separate behaviors
%     \item Acceleration data complex, we would possibly need more states with parametric model
%     \item Figure shows that summarise into 3 distinct behaviors that appear in batches, but for example, state 1 could not have been adequately modelled parametrically, sometimes more acceleration even in passive state (can be some sub-behavior we are not directly interested in but can invalidate inference.)
% \end{itemize}

\subsection{Spanish energy prices}
\label{subsec: energy}

For a practical demonstration of MS-GAM(LSS), we consider a time series of Spanish energy prices included in the R package \texttt{MSwM} \citep{sanchezMSwM2021}, and model the conditional distribution by an MS-GAMLSS similar to \citet{adam2022gradient}. The aim here is not to generate new insights or present an exhaustive analysis of the intricacies of the data but rather to demonstrate the practicality of qREML for smoothness selection in MS-GAMLSS. 

The data set comprises 1784 daily observations of energy prices (in Cents per kWh) in Spain which we want to explain using the daily oil prices (in Euros per barrel) also provided in the data. Specifically, we consider a 2-state MS-GAMLSS defined by
$$
\text{price}_t \mid \{ S_t = i \} \sim \mathcal{N} \bigl(\mu_t^{(i)}, (\sigma_t^{(i)})^2 \bigr),
$$
$$
\mu_t^{(i)} = \beta_{0,\mu}^{(i)} + s_{\mu}^{(i)}(\text{oil}_t), \quad \text{log}(\sigma_t^{(i)}) = \beta_{0, \sigma}^{(i)} + s_{\sigma}^{(i)}(\text{oil}_t), \quad i = 1,2,
$$
not covering other potential explanatory covariates for the sake of simplicity. We specify each smooth function as a linear combination of 12 cubic B-spline basis functions, with the associated design and penalty matrices provided by the R package \texttt{mgcv} \citep{wood2015package}.

\begin{figure}
    \centering
    \includegraphics[width=1\textwidth]{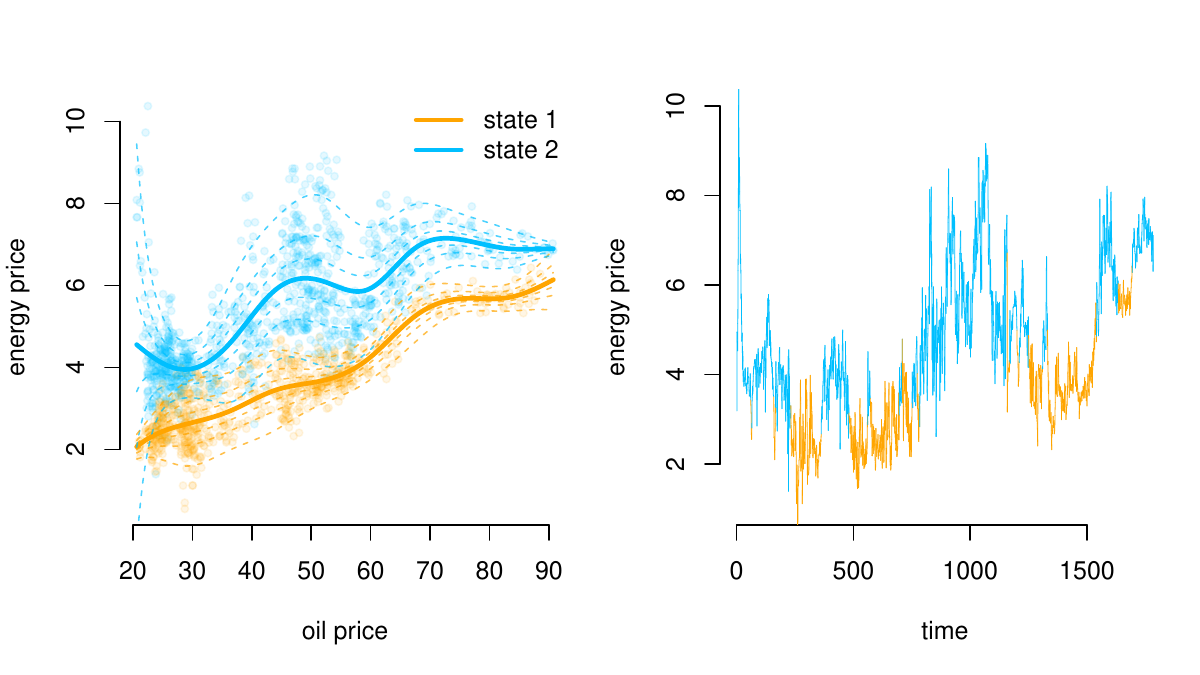}
    \caption{Estimated conditional means (solid lines) and quantiles (dashed lines) of the state-dependent distribution as a function of the oil price (left panel) and energy price time series colored according to the Viterbi-decoded state sequence (right panel).}
    \label{fig:energy}
\end{figure}

We fitted the model via qREML as detailed in Section \ref{sec:smoothness}, initializing the penalty strengths with $\bm{\lambda}_0 = (10^5, 10^5, 10^5, 10^5)$ and terminating the algorithm once the largest relative change of all penalty strengths fell below $10^{-4}$. Convergence was achieved after 14 iterations and model fitting took 7.65 seconds on an Apple M2 chip. For comparison, fitting the model via the full REML method with \texttt{RTMB} took 38.77 seconds and marginal maximum likelihood took 71.41 seconds. Applying a grid search method with five grid points would have resulted in $5^4 = 625$ models to be fitted --- each one potentially 10-20 times when using cross-validation.

The final penalty strengths were estimated as $\Hat{\bm{\lambda}} = (22.56, 7.21, 8.27, 4.17)$. Figure \ref{fig:energy_oil_modseq} in Appendix \ref{A:additional figures} shows how the estimation of the conditional means evolves over six iterations of the qREML algorithm. In particular, we see that the initial model is effectively a Markov-switching linear regression model, and that there is no visible difference in the fit anymore between iterations 5 and 10, thus a larger convergence threshold might have been sufficient in this case.

The off-diagonal entries of the t.p.m.\ were estimated as $\Hat{\gamma}_{12} = 0.019$ and $\Hat{\gamma}_{21} = 0.013$, indicating high persistence in the states with a mean state dwell time of 53 and 79 days in states 1 and 2 respectively.
The left panel in Figure \ref{fig:energy} shows the estimated conditional means in the two states as well as the 0.65, 0.80, and 0.95 percent quantile of the associated conditional normal distribution. The right panel in the same figure shows the time series colored according to the Viterbi-decoded state sequence in the right panel. 

Both states indicate a mostly positive effect of the oil price on the energy price which levels off for oil prices greater than 70 euros. The effect on the standard deviation differs strongly across the two states with a minor effect in the first state but a strong, non-linear effect in the second state, with volatility peaking for very low oil prices (< 20) and moderate oil prices (40-60). Overall, the standard deviation in the second state is substantially larger than in the first, indicating a link between the first state to a calm market regime and the second state corresponding to a nervous market regime.

\subsection{African elephant}

To illustrate smoothness selection for nonparametric modeling of transition probabilities within HMMs, we consider the movement track of an African elephant from the Ivory Coast, comprising 12170 GPS measurements of longitude and latitude, collected every two hours between September 2018 and November 2021. The data are available in the movebank repository 2736765655 \citep{movebank2024}. From the location data, we calculate step lengths and turning angles between consecutive observations, aiming to study the behavioral diel variation of the elephant. 

We model the data using a 2-state HMM, with gamma and von Mises state-dependent distributions for the step lengths and turning angles, respectively, assuming independence of the two variables conditional on the underlying behavioral state. As a baseline, we fit an HMM with homogeneous transition probabilities. Based on this model we can investigate the estimated state-dependent distributions (cf. Figure \ref{fig:marginal_elephant} in Appendix \ref{A:additional figures}), suggesting an interpretation of the two states as ``encamped'', characterized by relatively small step lengths and low persistence in turning angle, and ``exploratory'', characterized by larger step lengths and a stronger angular persistence.

\begin{figure}
    \centering
    \includegraphics[width=1\textwidth]{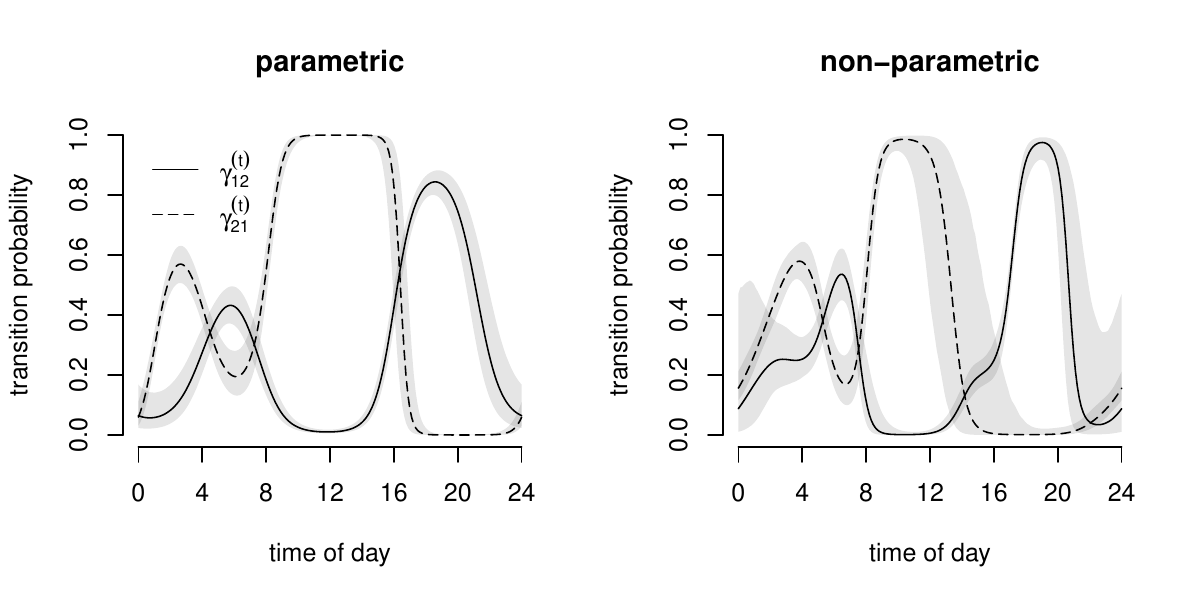}
    \caption{Transition probabilities as a function of the time of day of the parametric model (left panel) and
    nonparametric model (right panel). Pointwise 95\% confidence intervals (shown in gray) are obtained by sampling from the joint distribution of the MLE --- for the nonparameteric model as described in Section \ref{subsec:practical}.}
    \label{fig:transprobs_elephant}
\end{figure}

To characterize the behavioral variation as a function of the time of day, we model the transition probabilities nonparametrically using cyclic P-splines, which are wrapped at the boundary of the support to ensure diurnal continuity \citep{feldmann2023flexible}, leading to the model formulation
$$
    \text{logit}(\gamma_{ij}^{(t)}) = \beta_0^{(ij)} + s^{(ij)}(\text{time}_t), \quad i \neq j.
$$
Traditionally, such temporal dependencies have been modeled by performing a basis expansion into a small number of sine and cosine frequencies \citep{papastamatiou2018activity, beumer2020application} and for comparison we also fit such a parametric model with two sine and cosine frequencies, corresponding to a cycle length of 24 and 12 hours.

For practical implementation of the cyclic splines, we use the design and penalty matrix provided by the default option implemented in \texttt{mgcv} \citep{wood2015package} when setting \texttt{bs = 'cp'} (cyclic P-splines). We then fit the model by qREML, initializing the penalty strength with $\bm{\lambda}_0 = (10^5, 10^5)$ which produces an initial fit very close to a homogeneous HMM. Again using the default tolerance of $10^{-4}$, the algorithm converged in 11 iterations and the model fit took 11.83 seconds on an Apple M2 chip. In this case, the full REML method took 38.79 seconds and the full marginal ML method took 190.98 seconds while demonstrating issues with convergence.

Similarly to the previous case study, Figure \ref{fig:elephant_model_sequence} in Appendix \ref{A:additional figures} shows the estimated transition probabilities over the course of the iterations, demonstrating the smooth transition from the almost homogeneous model to the final, flexible fit. Again, convergence is very quickly achieved as there is no visual difference anymore between the fit in iterations 5 and 12. The final penalty strength was selected as $\Hat{\bm{\lambda}} = (0.248, 0.108)$, and the effective degrees of freedom for the two smooths are $8.18$ and $6.53$, respectively.

\begin{figure}
    \centering
    \includegraphics[width=0.88\textwidth]{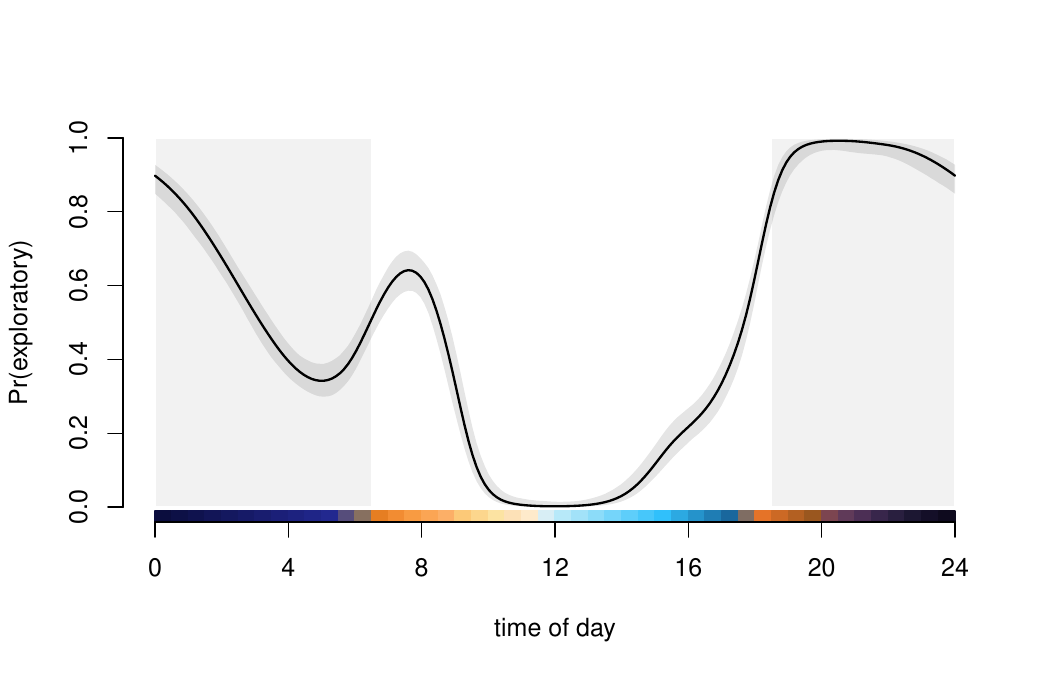}
    \caption{Periodically stationary distribution of the HMM fitted to the elephant data. Pointwise 95\% confidence intervals are shown light blue. The sun cycle is indicated at the bottom and the shaded gray areas indicate darkness.}
    \label{fig:stationary_elephant}
\end{figure}

Figure \ref{fig:transprobs_elephant} shows the estimated relationship between the transition probabilities and the time of day for the parametric model with two sine and cosine frequencies (left panel) and the nonparametric spline fit (right panel). While we see that the estimated relationship differs in shape due to the limited flexibility of the parametric model, we also find the following additional advantage of the nonparametric fit. For specific times of the day, some transitions are rarely present in the data ---
for example, around noon the elephant is almost always in the ``encamped'' state, such that the transition probability from ``exploratory'' to ``encamped'' cannot be estimated reliably during that time. Due to the local support of the basis functions, they account for the varying uncertainty across the covariate range as their coefficients are estimated with varying precision, ultimately leading to realistic fluctuation in the width of the pointwise confidence intervals. %, representing the time-varying uncertainty.
In contrast, the sine and cosine expansions have global support, hence the locally varying uncertainty cannot propagate to the function level and the width of the confidence intervals shows little variation.

Finally, Figure \ref{fig:stationary_elephant} shows the periodically stationary state distribution 
% derived from the transition probabilities 
introduced by \citet{koslik2023inference}. 
% which has the advantage of being much easier to interpret compared to the transition probabilities. 
We see that the animal has the highest probability of being active in the evening hours after 7 pm, which may be attributed to cooler temperatures, while being mostly inactive around noon, with a slight peak in immediately after sunrise.

\section{Discussion and Outlook}

To date, smoothness selection has been the main limitation for nonparametric Markov-switching models, primarily because for time series cross-validation is neither straightforward nor computationally efficient. Recently, \citet{michelot2022hmmtmb} proposed an alternative by treating the spline coefficients as random effects and integrating them out for model estimation using the \texttt{TMB} package. While this method is promising and seems more reliable than grid-search approaches, it remains computationally demanding even with the use of automatic differentiation.
%Moreover, its reliance on \texttt{TMB} necessitates model implementation in C++, which can be inconvenient for researchers who need to use custom models and wish to rapidly modify the code.

In this contribution, we introduced the \textit{qREML} algorithm, an efficient approximation of restricted maximum likelihood estimation (REML) for smoothness selection in these models. By exploiting the simple structure of the penalized likelihood, qREML enables semi-analytic derivations, resulting in substantial computational savings. Hence, qREML greatly simplifies and speeds up the estimation of
%provides a promising tool for estimating 
Markov-switching models with nonparametric components,
%In this contribution, we demonstrated that restricted maximum likelihood estimation (REML) is suitable for smoothness selection in such models. Notably, for spline smoothing, the structure of the penalized likelihood permits semi-analytic derivations, ultimately leading to substantial computational savings. Hence, we propose the qREML algorithm developed in Section \ref{sec:smoothness} as a promising tool for estimating Markov-switching models with nonparametric components, 
making the practical application of such models accessible to a wider audience.
% Our results indicate that REML not only simplifies the implementation process but also ensures more efficient and accurate model estimation. The semi-analytic derivations of the penalized likelihood in REML enable faster computations, significantly enhancing the feasibility of using nonparametric methods in complex models. 
This advancement is particularly relevant for models that require high computational resources.
%thereby broadening the scope of applications for nonparametric Markov-switching models. 
%We only implemented model estimation with gradients evaluated via finite differencing but implementing the method using automatic implementation will facilitate even faster model estimation.

The joint likelihood structure does in fact remain identical when considering independent and identically distributed random effects $\bm{b}_i \sim \mathcal{N}(0,\sigma_i^2 \bm{I})$ by setting $\lambda_i = 1 / \sigma_i^2$ and $\bm{S}_i = \bm{I}$. Such random effects are very popular in practice, for example in statistical ecology, where researchers often fit models to data comprising tracks of several animals, with the random effects either controlling for inter-individual variation or the estimate of $\sigma_i^2$ being of interest as the population variance. The practical potential of qREML for such kinds of models needs further evaluation as the estimation accuracy for $\sigma_i^2$ should be higher compared to spline smoothing where the estimated penalty strength is only of minor interest.
%Initial experiments are promising, suggesting that this method could enhance the efficiency of Markov-switching models including both penalized splines and simple random effects.

Given the very general setting considered in Section \ref{sec:smoothness}, it is evident that the smoothness selection procedure is not limited to the specific types of Markov-switching models discussed in this contribution, but instead can be extended to several closely related classes of models, including, inter alia, continuous-time HMMs, state-space models and Markov-modulated Poisson processes \citep{mews2024build}. In any of these classes of models, the latent nature of the state process makes nonparametric inference particularly desirable and valuable, as the inability to conduct exploratory data analysis separately for each of the model components --- such as the distinct emission distributions or covariate effects on different transition probabilities --- renders the formulation of adequate parametric models challenging. This contribution is a major step in improving the feasibility of any corresponding nonparametric analysis, which will hopefully lead to an increased uptake of these methods by practitioners.

\section*{Supplementary materials}
The data and code for fully reproducing all case studies and the simulation experiments can be found at \url{https://github.com/janoleko/qREML}. The code for the qREML procedure can be found at \url{https://github.com/janoleko/LaMa/blob/main/R/qreml_functions.R}

\section*{Acknowledgments}

The author is very grateful to Katie Harrington for providing the caracara data and sincerely thanks Roland Langrock, Dietmar Bauer and Thomas Kneib for their helpful comments on an earlier version of this manuscript.

%Bibliography
% \bibliographystyle{unsrt}  
% \bibliography{references}  

\printbibliography

\newpage

\appendix
\renewcommand{\thesubsection}{\Alph{section}.\arabic{subsection}}

\section{Appendix}

\subsection{Simulation experiments}

\label{sec:simulation}

The following simulations are primarily supposed to demonstrate the general practicality of the qREML approach. In addition, we will investigate the number of iterations needed and its dependence on the sample size.
% We only consider the scenario of transition probabilities smoothly depending on external covariates, as this is the most challenging of the three model formulations from our perspective because transition probabilities operate regarding the latent part of the model, hence not being directly observed. Additionally, smoothness selection for this aspect of the model is particularly influenced by the time-series nature of the data, making cross-validation approaches challenging and not clear apriori.
We focus exclusively on the scenario where transition probabilities vary smoothly with covariates. This approach may be considered the most challenging of the three model formulations described in Section \ref{sec:model formulation}, primarily because state transitions pertain to the latent component of the model and are therefore not directly observed. The difficulty is compounded by the time-series nature of the data, making traditional cross-validation techniques problematic and their effectiveness unclear in advance.

The data are generated by a 2-state Gaussian HMM with state-dependent means $\bm{\mu} = (1, 5)$ and standard deviations $\bm{\sigma} = (1, 3)$. The transition dynamics are governed by transition probabilities that depend on a covariate $z_t$:
$$\text{logit}(\gamma_{12}^{(t)}) = -2 + \sin(3 \pi z_t) + \exp(1.5 z_t), \quad \text{logit}(\gamma_{21}^{(t)}) = 2 + \cos(4 \pi z_t) - 2 \exp(z_t),$$
where $z_t$ is drawn independently from a uniform distribution on $(0,1)$.

For varying time-series lengths of $T = 1000, 2000, 5000$, we simulated 200 data sets each and fitted a nonparametric HMM via qREML, as discussed in the Section \ref{sec:smoothness}. Specifically, the two off-diagonal transition probabilities are modeled smoothly using the B-spline design and penalty matrix as provided by \texttt{mgcv} when specifying \texttt{bs = `ps'}, with 15 basis functions to allow for sufficient flexibility. In each simulation run, the penalty strength vector is initialized as $\bm{\lambda}_0 = (1000, 1000)$. For the three time-series lengths, the average estimation times per simulation run were 3.12, 4.16, and 9.66 seconds, respectively, on an Apple M2 chip.

\begin{figure}
    \centering
    \includegraphics[width=1\textwidth]{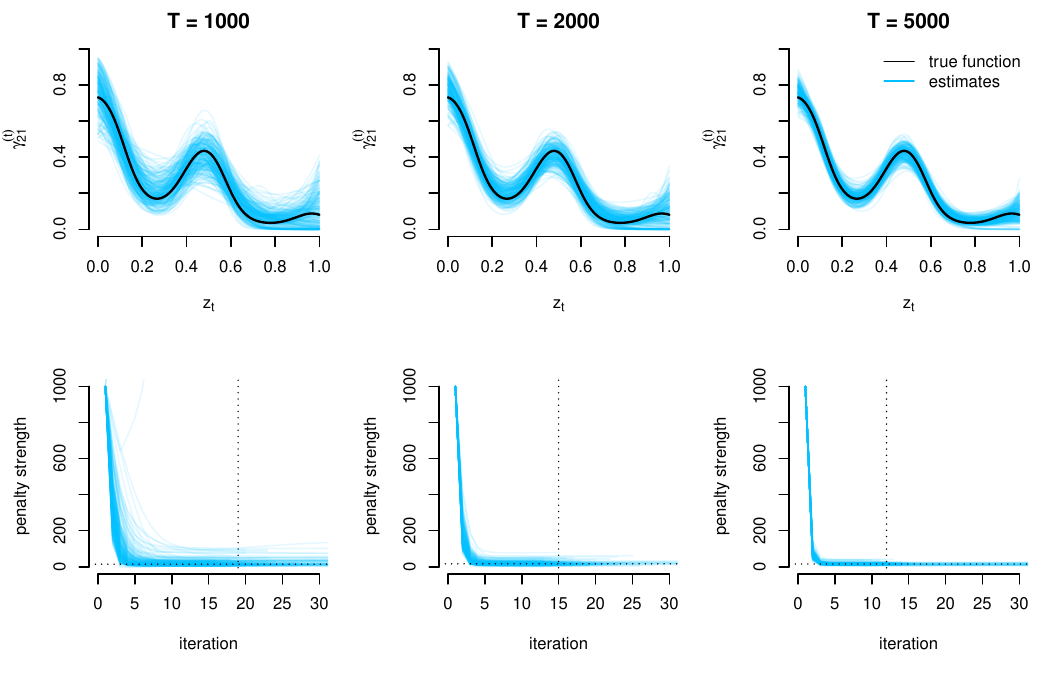}
    \caption{Top panel: Estimated transition probabilities $\gamma_{21}^{(t)}$ (light blue) obtained from simulating 200 data sets from the specified HMM, for varying amounts of data and true relationship (black). Bottom panel: trace plots of the penalty strength for $\gamma_{21}^{(t)}$ (light blue lines) complemented with the median iterations needed for convergence (vertical dotted line) and median penalty strength at the last iteration (horizontal dashed line).}
    \label{fig:sim_transprobs2}
\end{figure}

Figure \ref{fig:sim_transprobs2} displays the results for the three simulation scenarios with varying $T$. The top three panels show the smooth fits for $\gamma_{21}^{(t)}$ as a function of $z_t$, whereas the bottom three panels display trace plots of the associated penalty strength for each simulation run. The results for $\gamma_{12}^{(t)}$ are shown in Figure \ref{fig:sim_transprobs} in Appendix \ref{A:additional figures}.
Based on visual inspection of the overall results, we find that the smoothness penalty was chosen adequately in almost all runs, reliably producing satisfactory function estimates.
% Overall, the results are very promising, as there seems to be no visual bias and the smoothness of the estimated functions is chosen adequately. 
Unsurprisingly, the variance of the estimated function decreases for increasing $T$, hence we conclude that modeling the transition probabilities as smooth functions of external covariates requires a considerable amount of data. 

Notably, the convergence of the qREML algorithm becomes substantially faster and more reliable for increasing $T$. 
For $T=1000$ 
%we see a notable amount of models that seemed to converge to an inadequate penalty strength and 
convergence at a tolerance of $10^{-5}$ took a median of $\sim 17$ iterations, with many models needing more than 25 iterations. For %$T=2000$ and $T=5000$, the performance increased drastically, and for 
$T=5000$, the vast majority of models converged in 15 iterations or less, with a median of $\sim 12$ iterations until convergence. 
This can be explained by the likelihood function more and more resembling the shape of a multivariate Gaussian for larger $T$ leading to the Laplace approximation being more accurate \citep{potscher2013dynamic}.
%Hence, this conclusion should also hold similarly for the full REML and full marginal ML methods.

\newpage

\subsection{Additional figures}
\label{A:additional figures}

\begin{figure}[h]
    \centering
    \includegraphics[width=1\textwidth]{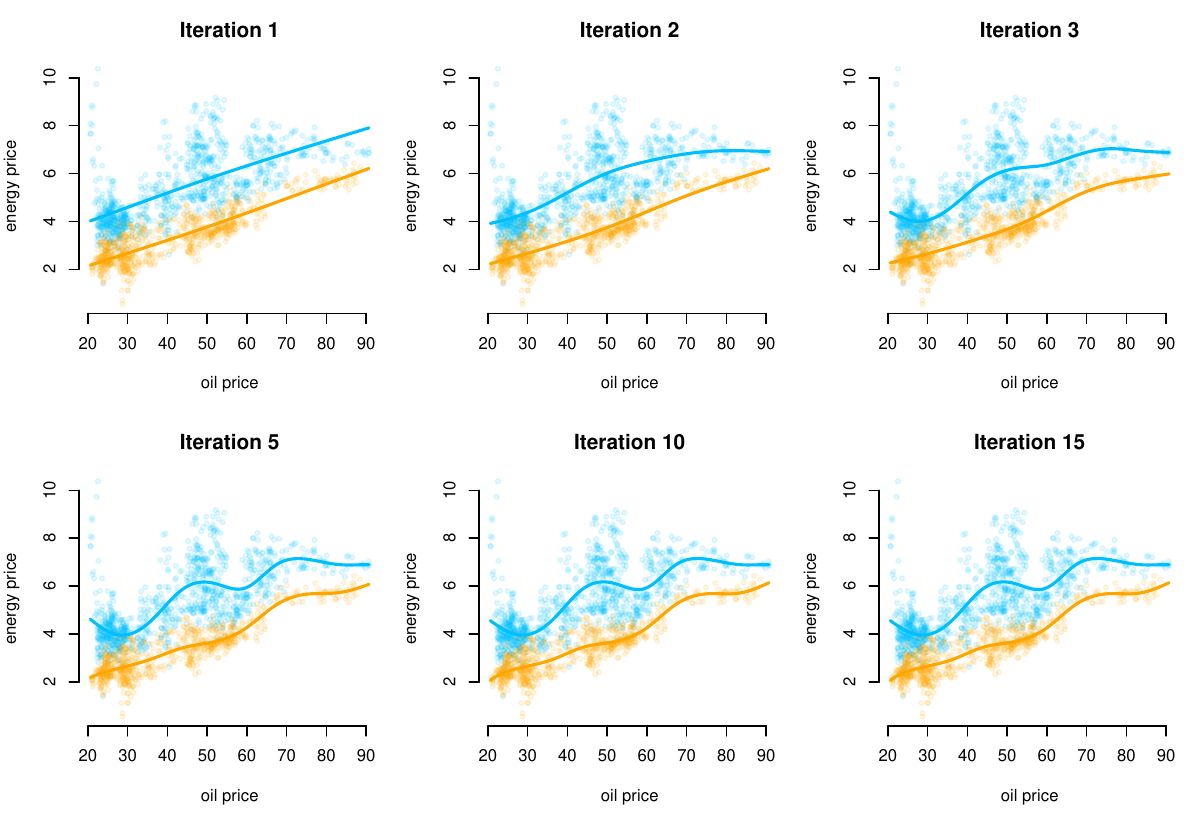}
    \caption{Estimated MS-GAMs for different iterations of the qREML fit.}
    \label{fig:energy_oil_modseq}
\end{figure}

\begin{figure}
    \centering
    \includegraphics[width=1
\textwidth]{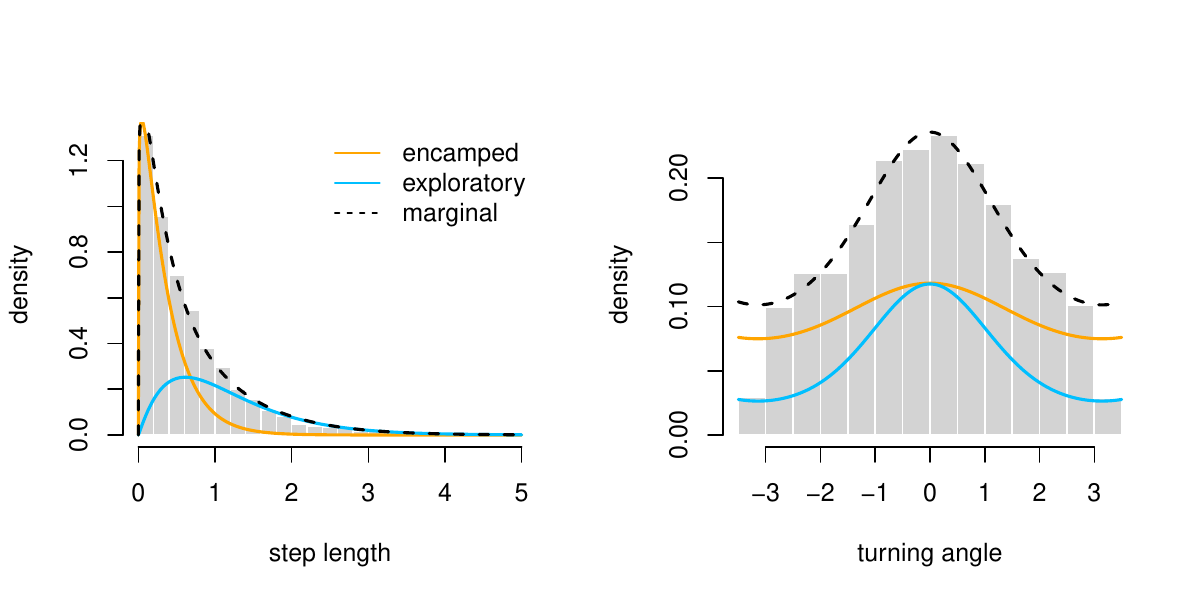}
    \caption{Weighted state-dependent step-length (left panel) and turning angle (right panel) distributions in the encamped (orange) and exploratory (light-blue) state, complemented with the marginal distribution (black)}
    \label{fig:marginal_elephant}
\end{figure}

\begin{figure}
    \centering
    \includegraphics[width=1\textwidth]{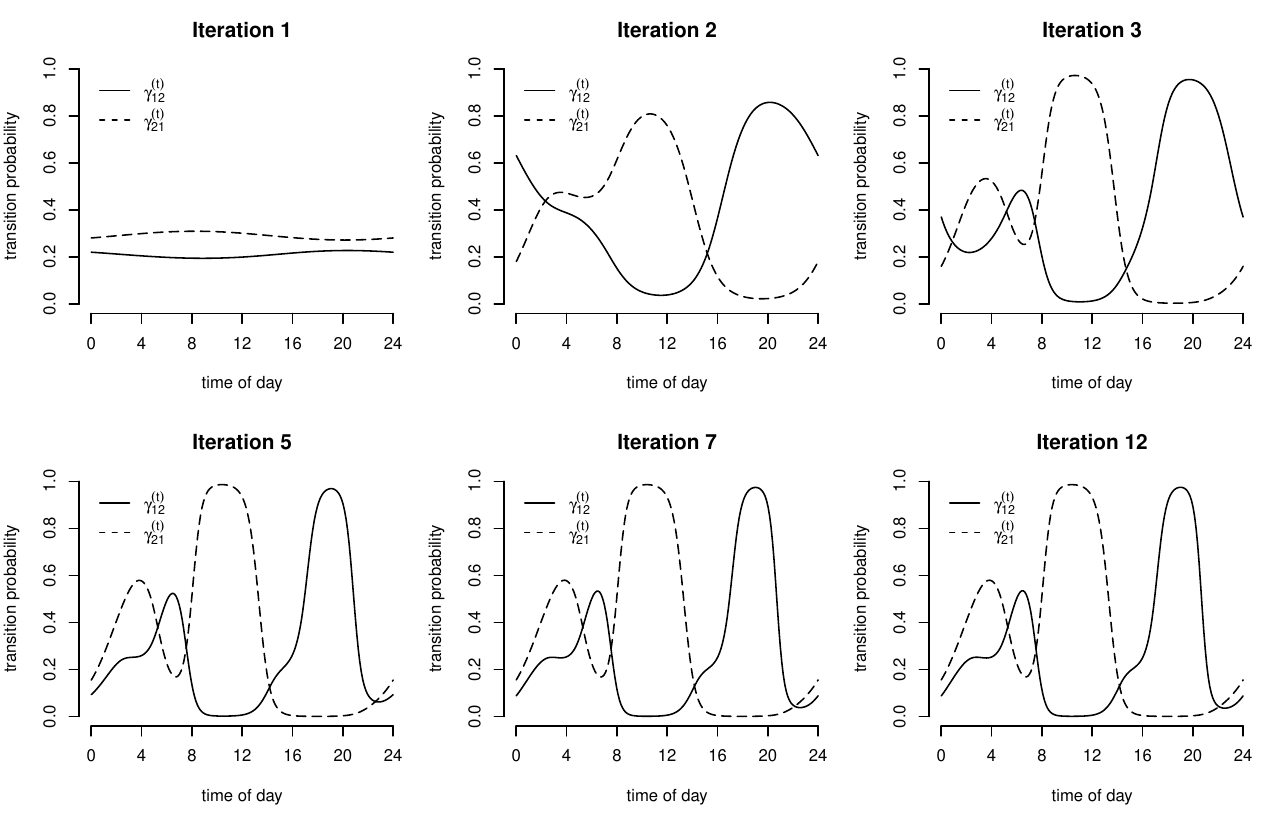}
    \caption{Transition probabilities as a function of the time of day, derived from different iterations of the qREML fit.}
    \label{fig:elephant_model_sequence}
\end{figure}

\begin{figure}
    \centering
    \includegraphics[width=1\textwidth]{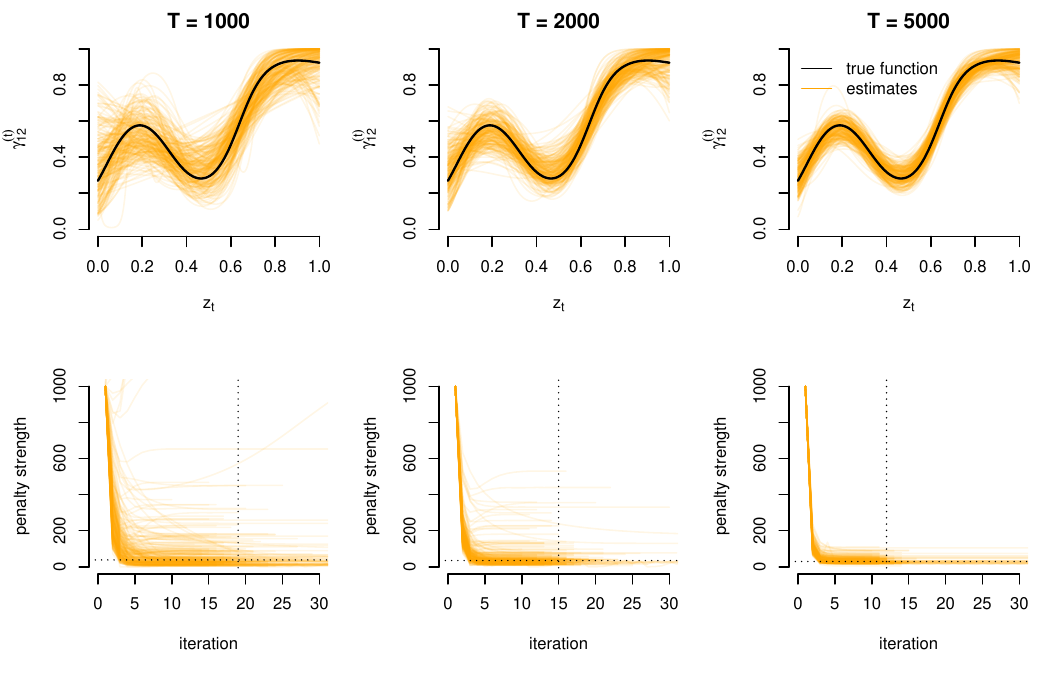}
    \caption{Top panel: Estimated transition probabilities $\gamma_{12}^{(t)}$ (orange) obtained from simulating 200 data sets from the specified HMM, for varying amounts of data and true relationship (black). Bottom panel: trace plots of the penalty strength for $\gamma_{12}^{(t)}$ (orange) complemented with the median iterations needed for convergence (vertical dotted line) and median penalty strength at the last iteration (horizontal dashed line).}
    \label{fig:sim_transprobs}
\end{figure}

\end{document}